\documentclass[12pt]{article}
\newcommand{\be}{\begin{equation}}
\newcommand{\ee}{ \end{equation}}
\newcommand{\ba}{\begin{array}}
\newcommand{\ea}{\end{array}}

\newcommand{\cW}{ {\cal W}}
\newcommand{\cT}{ {\cal T}}
\newcommand{\cU}{ {\cal U}}
\newcommand{\cV}{ {\cal V}}
\newcommand{\hW}{ h_{\cal W}}
\newcommand{\hT}{ h_{\cal T}}
\newcommand{\hU}{ h_{\cal U}}
\newcommand{\qW}{ q_{\cW}}
\newcommand{\qT}{ q_{\cT}}
\newcommand{\qU}{ q_{\cU}}

\let\LARGE=\Large
\let\Large=\large

 \def\unit{\hbox to 3.3pt{\hskip1.3pt \vrule height 7pt width .4pt \hskip.7pt
\vrule height 7.85pt width .4pt \kern-2.4pt
\hrulefill \kern-3pt
\raise 4pt\hbox{\char'40}}}

\begin{document}


\thispagestyle{empty}
\rightline{DAMTP-98-76}
\rightline{IC/98/66}
\rightline{QMW-PH-98-29}
\rightline{hep-th/9807014}

\vspace{1truecm}

\centerline{\bf \LARGE  Black Holes and Flop Transitions}

\vspace{0,4cm}

\centerline{\bf \LARGE in M-Theory on Calabi-Yau Threefolds }

\vspace{1truecm}

\centerline{
{\bf Ingo Gaida}$^{a,\#}$, {\bf Swapna Mahapatra}$^{b, *}$,
{\bf Thomas Mohaupt}$^{c}$, {\bf Wafic
A. Sabra}$^{d}$\footnote{E-mail: 
I.W.Gaida@damtp.cam.ac.uk, swapna@beta.iopb.stpbh.soft.net, 
\\ 
mohaupt@hera1.physik.uni-halle.de, W.Sabra@qmw.ac.uk.
\\
$^{\#}$Research supported by Deutsche Forschungsgemeinschaft (DFG),
\\
$^*$Regular Associate of the ICTP, 
Permanent address: 
Physics Department, Utkal University,\\
Bhubaneswar-751004, India.
}}

\vspace{.5truecm}

\centerline{\em $^a$ DAMTP, University of Cambridge,
Silver Street, Cambridge CB3 9EW, UK}
\vspace{.2truecm}
\centerline{\em $^b$ Abdus Salam International Centre for 
Theoretical Physics, 34100 Trieste, Italy}
\vspace{.2truecm}
\centerline{\em $~^c$ Martin-Luther Universit\"at, Fachbereich Physik,
06099 Halle, Germany}
\vspace{.2truecm}
\centerline{\em $^d$ Physics Department, Queen Mary and Westfield 
College, Mile End Road, E1 4NS}



\vspace{.5truecm}

\begin{abstract}

\noindent We present fivedimensional extreme black hole solutions
of M-theory compactified on Calabi-Yau threefolds and study
these solutions in the context of flop transitions in the 
extended K\"ahler cone. 
In particular we consider a specific
model and present black hole solutions,
breaking half of $N=2$ supersymmetry, in two regions
of the extended K\"ahler cone, which are connected by a flop transition.
The 
conditions necessary to match both solutions at the flop
transition are analysed. Finally we also discuss the conditions to obtain
massless black holes at the flop transition.

\vspace{.5truecm}

\end{abstract}


\newpage


\section{Introduction}

In recent times there has been a lot of interest in the 
study of various 
dualities in string theory. The central 
observation is that the five distinct ten-dimensional
string theories can be 
thought of as the weak coupling limits of various compactifications
of elevendimensional M-theory, whose low-energy limit is given
by  elevendimensional supergravity \cite{Witten1}. This can 
be used to lift known dualities by one dimension.
In four dimensions $N = 2$ heterotic 
string theory on $K_3 \times T^2$ and type IIA string theory
on Calabi-Yau threefolds
are dual to each other \cite{Vafa}. Using M-theory
this duality can be lifted to five dimensions 
so that heterotic string theory on $K_3 \times
S^1$ is related to M-theory on Calabi-Yau spaces with $N=2$ 
supersymmetry \cite{Cadavid}. 
This is equivalent to taking the large volume limit of 
type II string theory compactified on the same Calabi-Yau manifold
\cite{Taylor}. 
\\
When compactifying M-theory on a Calabi-Yau manifold, the two-brane
and the five-brane can be wrapped around the two- and four-cycles of the 
Calabi-Yau space to give rise to BPS states of the 
$D = 5$ theory which has 
$N = 2$ supersymmetry \cite{Hull}. The study of BPS black hole solutions 
in the compactified theories has become interesting in the context
of a microscopic derivation of the macroscopic 
Bekenstein-Hawking entropy \cite{Hawking} through 
the counting 
of microscopic degrees of freedom of D-branes \cite{Strominger1}.
Moreover BPS solutions account for the 
additional light states that have to be present in certain regions
of moduli space in order to cure apparent singularities of the
low energy effective action and to describe phase transitions between
different branches of the moduli space \cite{Strominger}.
\\
BPS saturated solutions in toroidal compactifications have $N = 4$
and $N = 8$ supersymmetry and they do not receive quantum corrections.
On the other hand, BPS saturated solutions in $N = 2$ supergravity 
can receive quantum corrections at one-loop level. 
These $N = 2$ models 
with vector and hyper multiplets in four and five dimensions 
can arise as compactifications of type-II or M-theory on a 
Calabi-Yau threefold, respectively. The Bekenstein-Hawking entropy of 
these $N = 2$ black holes or 
more generally solitons with non-singular 
horizons can be obtained by extremizing the underlying central charge 
\cite{FerKal1} and the scalar fields take fixpoint values 
at the horizon, 
which are independent of their values at infinity. These fixed values
are determined in terms of conserved charges and topological 
data of the compactified space. The most simple 
example is the
double-extreme black hole, where the scalar fields 
(moduli) are constant throughout the entire space-time \cite{Double}. 
\\
Extreme 
black hole solutions with non-constant scalar fields have also 
been obtained in four \cite{Wafic1} 
and five dimensions \cite{Wafic}. 
Whereas the relevant four dimensional theory
is described by special geometry, 
in five dimensions the coupling of $N =2$ supergravity to abelian 
vector multiplets is based on the structure of very special geometry
\cite{Sierra,Proeyen}. 
As already mentioned above, moduli spaces of string compactifications 
have critical points where additional massless states appear. Sometimes
this is related to topology changing phase transitions of the internal
compact space. The moduli spaces relevant in the context of supersymmetric
black hole solutions with finite horizons are the K\"ahler moduli
spaces of four- and fivedimensional theories with $N\geq 2$ supersymmetry.
In four dimensions, there can be phase transitions between 
geometrical phases with a sigma model description and non-geometrical
phases corresponding to more abstract conformal field theories like 
Landau-Ginzburg models. In five dimensions, on the other hand,
it has been shown that all the phases of the M-theory compactification 
are geometrical and the phase transitions
are sharp \cite{Witten2}. 

Generically, topological phase transitions among $N=2$ string vacua
occur at points in the moduli spaces 
where the non-perturbative BPS states become massless. The well-known
example in four dimensions are the massless electrically or magnetically 
charged black holes at the conifold points in the Calabi-Yau moduli
spaces, where certain homology cycles of the Calabi-Yau spaces shrink 
to zero size \cite{Strominger}. The other examples are enhancement 
of gauge symmetry \cite{Witten3} and appearance of tensionless strings
\cite{Seiberg}. The interplay between Calabi-Yau
phase transition and the behaviour of black hole like space
time configurations has been studied in \cite{BehLueSab} in the
context of fourdimensional type II string compactifications. In this
paper we will perform a similar analysis in fivedimensional M-theory,
taking advantage of the fact that here various simplifications 
occur.
Namely, in the fivedimensional theory resulting from 
the compactification of M-theory on Calabi-Yau threefold, there
is no analogue of stringy $\alpha'$ corrections. In addition, there 
are no axionic fields and the moduli (the sizes of the 
two-cycles in 
the Calabi-Yau space) take values in a real K\"ahler cone. 
Unlike 
the four dimensional case, the transitions between various phases in 
five dimensions have to go through true singularities in 
the Calabi-Yau space and one obtains sharp phase transitions. 
There is 
a variety of possible phenomena that can occur as one approaches the boundary
of the K\"ahler cone. The so called "flop transition" is one such 
example, where one of the moduli approaches the boundary of its
original range of values and subsequently 
is analytically continued to a new region. 
Geometrically this corresponds to a transition into a new K\"ahler cone 
which is assigned to a birationally equivalent Calabi-Yau with 
different intersection numbers, but the same Hodge numbers 
\cite{Witten2}. 

A study of critical points and phase transitions in five dimensions
in the case of double extreme black holes has been presented
in \cite{Rahmfeld}.
In this article we study these phase transitions
in the context of extreme black hole solutions in five dimensions, 
i.e. we construct and analyze solutions that break half of $N=2$ 
supersymmetry, with scalar fields which are not constant throughout 
the entire space-time.
Thus, the scalars represent dynamical degrees of freedom of the
model. 

The paper is organized as follows. In sections \ref{SecSuGra} and
\ref{SecM} 
we give a brief 
review of $D=5$, $N=2$ supergravity and very special geometry 
and its connection to M-theory compactified on a Calabi-Yau 
threefold, respectively. 
In section \ref{SecSol}, we analyze the static, extreme black
hole solutions in five dimensions. In section \ref{SecMod}, we 
present the so called ${\bf F}_1$ model and the prepotentials 
of its two K\"ahler cones, which are connected by a flop transition. 
In sections \ref{Sec2} and \ref{Sec3} we present the black hole solutions 
in both 
regions, namely region II and III \cite{Rahmfeld}, by solving 
the stabilization equations
and we 
obtain the ADM mass and the 
entropy of the black hole by evaluating 
the central charge at infinity and at the horizon, respectively.
In section \ref{SecInter}, 
we consider generic supersymmetric black hole solutions
in five dimensions, which interpolate between flat space at 
$r = \infty$ and a Bertotti-Robinson geometry at the horizon at 
$r = 0$. We analyze the behaviour of solutions when the
moduli at infinity and at the horizon take values 
in different regions (K\"ahler cones).
In section \ref{SecMassless}, 
we find the conditions for obtaining massless 
black hole solutions at the flop transition. Finally we 
summarize our results in section \ref{SecCon}.

\section{D=5, N=2 Supergravity and Very Special Geometry \label{SecSuGra}}

The action of five dimensional $N=2$ supergravity coupled to 
$N=2$ vector multiplets has been constructed
in \cite{Sierra}.
In the following we will consider
compactifications of $N=1$, $D=11$ supergravity, i.e. the low
energy limit of M-theory,
down to five dimensions on Calabi-Yau 3-folds ($CY_{3}$) with
Hodge numbers ($h_{(1,1)},h_{(2,1)}$) and topological intersection numbers
$C_{IJK} (I, J, K = 1, \ldots , h_{(1, 1)})$. The five dimensional 
theory contains the gravity multiplet $(e_{A\mu}, \Psi_{\mu I}, 
{\cal A}_{\mu})$ \ \ \ $(I = 1, 2)$, $h_{(1, 1)} - 1$ vector multiplets
$({\cal A}_{\mu}^A, \lambda_I^A, \phi^A)$  \ \ $(A = 1, \ldots , 
h_{(1, 1)} - 1)$ and $h_{(2, 1)} + 1$ hypermultiplets $(\zeta^m, 
{\cal A}_I^m)$ \ \ $(m = 1, \ldots, 2(h_{(2, 1)} + 1))$. 
The $N_V$-dimensional space $\cal M$ ($N_V=h_{(1,1)}-1$) of scalar 
components of $N=2$ abelian vector multiplets coupled to supergravity
can be regarded as a hypersurface of a $h_{(1,1)}$-dimensional manifold
whose coordinates $X(\phi)$ are in correspondence with the vector bosons
(including the graviphoton). The definining equation of the
hypersurface is 
\begin{eqnarray}
{\cal V}(X) &=& 1 
\end{eqnarray}
and the prepotential ${\cal V}$ is a homogeneous cubic polynomial
in the coordinates $X(\phi)$:
\begin{eqnarray}
{\cal V}(X) &=& \frac{1}{6} \ C_{IJK}
                X^I X^J X^K,
\hspace{1cm} I,J,K = 1, \ldots h_{1,1} 
\end{eqnarray}
In five dimensions the $N=2$ vector multiplet has a single scalar and
$\cal M$ is therefore real. 
The special case corresponding to
perturbative heterotic compactifications on $K_3 \times S^1$ has
factorizable prepotential 
\begin{eqnarray}
{\cal V}(X) &=& X^1 \ Q(X^{I+1}),
\hspace{1cm} I = 1, \ldots N_V 
\end{eqnarray}
where $Q$ denotes a quadratic form. It follows that the scalar 
fields parameterize the coset space
\begin{eqnarray}
\label{space}
 {\cal M} &=& SO(1,1) \ \times \ \frac{SO(N_V-1,1)}{SO(N_V-1)}. 
\end{eqnarray}
The bosonic action of fivedimensional $N=2$ supergravity coupled to 
$N_V$ vector multiplets is given by 
(with metric $(-,+,+,+,+)$ and omitting Lorentz indices)
\begin{eqnarray}
e^{-1} {\cal L} = - \frac{1}{2} R
                     - \frac{1}{2} g_{ij} \partial\phi^i \partial\phi^j
                     - \frac{1}{4} G_{IJ} 
                          F^I F^J
                     + \frac{e^{-1}}{48} C_{KLM} 
                          \epsilon F^K F^L A^M.
\end{eqnarray}
The corresponding vector and scalar metrics are completely
encoded in the function ${\cal V}(X)$ 
\begin{eqnarray}
 G_{IJ} &=& - \frac{1}{2} \frac{\partial}{\partial X^I} 
                          \frac{\partial}{\partial X^J} 
                       \ln {\cal V}(X)_{|{\cal V}=1},
\\
 g_{ij} &=& G_{IJ} \frac{\partial}{\partial \phi^i} X^I(\phi)
                   \frac{\partial}{\partial \phi^j} X^J (\phi)_{|{\cal V}=1}.
\end{eqnarray}
Moreover it is convenient to introduce ``dual'' special coordinates
\begin{eqnarray}
 X_I \equiv \frac{1}{6} C_{IJK} X^J X^K
\ \ \Rightarrow \ \ X^I X_I = 1.
\end{eqnarray}

\section{Connection of D=5, N=2 Supergravity to M-theory and 
String Theory \label{SecM}}

As already mentioned above we consider $D=5$, $N=2$ supergravity as a
compactification of elevendimensional supergravity on a 
Calabi-Yau threefold ($CY_3$) \cite{Cadavid}.
The corresponding non-perturbative definition of 
elevendimensional supergravity is provided by M-theory. 
Alternatively one can 
interpret this setup as type IIA string theory compactified to four 
dimensions
on the same $CY_3$ in the limit of large K\"ahler structure
and large type IIA coupling \cite{Taylor, Witten2}. 
In this limit, $CY_3$ completely
decompactifies with respect to the IIA metric. However, in terms of
M-theory variables the volume of $CY_3$ is kept fixed 
whereas the M-theory circle decompactifies.
\\
Going from $D=4$ to $D=5$, $N=2$ supergravity results 
in various simplifications in the
geometry of the K\"ahler moduli space and hence in the dynamics of
vector multiplets. 
Since the key features are important in what follows, we briefly
review them, following \cite{Witten2}:

(i) The M-theory limit of the IIA K\"ahler
moduli space can be thought of as a zero slope limit in which stringy
effects, i.e. $\alpha'$ corrections, are switched off. As a consequence
the fivedimensional prepotential is purely cubic.
The coefficient of the cubic term is given by the triple intersection
numbers $C_{IJK}$ of homology four-cycles. 

(ii) The K\"ahler moduli in four dimensions are complex due to 
the presence of axion-like scalars, whereas they
are real in five dimensions. Thus, the K\"ahler moduli space of
the fivedimensional theory is the standard real K\"ahler cone 
of classical geometry, in contrast to the complexified
``quantum'' K\"ahler moduli space of string compactifications.\footnote{
See for example \cite{GreeneRev} for a review.}
Finally
the modulus corresponding to the total volume of the Calabi-Yau space
sits in a hypermultiplet of the fivedimensional theory. 

Concerning the global structure of the K\"ahler moduli space, the
M-theory limit has the effect of eliminating the non-geometric phases
encountered in four dimensions, because these are due to $\alpha'$
effects. On the other hand the transition between different
geometric phases is still possible through so-called
``flop transitions'' \cite{Witten2}.\footnote{
We do not consider phase transitions that involve tuning the 
hypermultiplets, because the solutions 
we study only depend on the vector multiplets. 
Transitions involving hypermultiplets have been studied extensively
in F-theory \cite{MorVaf, Louis}.} 
At a flop transition, one or several complex
curves in the manifold degenerate to zero volume. After transforming
the triple intersection numbers in a specific way depending on 
the particular degeneration one can continue into the K\"ahler cone
of another Calabi-Yau space. Gluing together the K\"ahler cones of
all Calabi-Yau spaces that are related by flop transitions, one obtains
the so-called ``extended K\"ahler cone''.\footnote{This is sometimes called
the partially extended K\"ahler cone, thus reserving the term 
extended K\"ahler cone to the case of stringy moduli spaces which 
include non-geometric phases.}
\\
In physical terms
one has additional massless hypermultiplets at the transition point
which descend from M-theory branes wrapped on the vanishing cycles.
Integrating these extra states out amounts to changing the triple
intersection numbers in precisely the way predicted by the Calabi-Yau
geometry. It will be important later that the low energy effective
action is regular at the transition. In comparison to fourdimensional
type II compactifications we note that there flop transitions are
``washed out'' by $\alpha'$ corrections and by the fact that the K\"ahler
moduli are complexified. This has the consequence that one can 
smoothly interpolate between flopped Calabi-Yau spaces in fourdimensional
string theory, whereas in fivedimensional M-theory the transition is sharp
with singular geometry, but regular physics. 
\\
Finally, the K\"ahler moduli spaces of fivedimensional
M-theory have boundaries where a further continuation of the moduli
is impossible. These boundaries are either related to 
the presence of additional massless states or to 
tensionless strings \cite{Seiberg}. In the first 
case the effective action is regular, whereas it is singular in the second one.
\\
Thus, the maximally extended K\"ahler moduli space in the M-theory 
limit is given by a hypersurface in
the (partially) extended K\"ahler cone, which contains all Calabi-Yau spaces
that can be related to one another by flop transitions.
Since no stringy $\alpha'$ corrections are present and since the
prepotential is simply a cubic polynomial, we can compute exact 
solutions to the low-energy effective action and analyze them in
detail, whereas in four dimensions one needs to expand around
special points in moduli space, as was done in \cite{BehLueSab}.
\section{Extreme Black Hole Solutions \label{SecSol}}
In the following we will analyse spherical symmetric
electrically charged BPS black hole
solutions breaking half of $N=2$ supersymmetry. We will use the
approach and general solution presented in \cite{Wafic}.
The black hole solution is given by the following set of
equations
\begin{eqnarray}
ds^2 &=& - e^{-4V(r)} \ dt^2 \ + \ e^{2V(r)} (dr^2 + r^2 \
d\Omega_3^2),
\nonumber \\
F^I_{tm} &=& - \partial_m(e^{-2 V}X^I), \ \ \ m=1,2,3,4,
\nonumber \\
3 e^{2V} X_I &=&  H_I \ \equiv \ h_I + \frac{q_I}{r^2}.
\end{eqnarray}
where, $F^I_{tm}$ is the electric field strength. Near the horizon 
of the black hole the metric function $e^{2V}$,
which is in general a function of harmonic functions,
satisfies
\begin{eqnarray}
\lim_{r \rightarrow 0} \ e^{2V} &=& \frac{1}{3} \ \frac{Z_h}{r^2},
\end{eqnarray}
where $Z=X^I q_I$ is the (electric) central charge,
appearing in the supersymmetry algebra, and $Z_h$ is its value at the horizon.
It follows that the Bekenstein-Hawking entropy \cite{Hawking} 
of extreme black holes in five dimensions
is given by \cite{FerKal1}
\begin{eqnarray}
S_{BH} &=& \frac{A}{4G_N} \ = \ 
           \frac{\pi^2}{2 G_N}\ \big |\frac{Z_h}{3} \big |^{3/2}.
\end{eqnarray}
Moreover, the ADM mass of the black hole solution is determined by the
central charge evaluated at spatial infinity:
\begin{eqnarray}
M_{ADM} &=& \frac{\pi}{4G_N} \ Z_{\infty}.
\end{eqnarray}
It is useful to introduce rescaled special coordinates
\begin{eqnarray}
 Y^I &=& e^{V} X^I, \ \ \ Y_I \ = \ e^{2V} X_I  
\end{eqnarray}
such that ${\cal V}(Y) = Y_I Y^I = e^{3V}$. It follows that the
background metric of the black hole solution is given by
\begin{eqnarray}
ds^2 &=& - {\cal V}^{-4/3}(Y) \ dt^2 \ + \ {\cal V}^{2/3}(Y) (dr^2 + r^2 \
d\Omega_3^2).
\end{eqnarray}
Moreover the stabilisation equations take the simple form
\begin{eqnarray}
C_{IJK} Y^J Y^K &=& 2 \ H_I. 
\label{StabEq}
\end{eqnarray}
Thus, near the horizon of the black hole one obtains
\begin{eqnarray}
\lim_{r \rightarrow 0} \ {\cal V}(Y) &=& 
\left (
  \frac{1}{3} \ \frac{Z_h}{r^2}
\right )^{3/2}
\end{eqnarray}
\section{The Model \label{SecMod}}
In the following we will restrict ourselves to 
one family of Calabi-Yau spaces. 
The reason is that fivedimensional black holes
have a complicated model dependence, because
(\ref{StabEq}) is a 
system of coupled quadratic equations, which can only be solved
on a case by case basis.
Thus solutions are determined
in terms of harmonic functions, 
but the way the metric and scalars depend on the harmonic functions
is much more complicated as it is, for instance,
for fourdimensional axion-free
black holes. 
\\
The particular model we choose was introduced in the context of
F-theory \cite{MorVaf}. We refer to it as the ${\bf F}_1$ model, since one 
of its phases is an elliptic fibration over the
first Hirzebruch surface ${\bf F}_1$.  
\\
We will use the notation introduced in \cite{Rahmfeld}. The 
(partially) extended K\"ahler cone can be covered by one set of
moduli (special coordinates) 
$X^{1,2,3}= (T,U,W)$.\footnote{Eventually one has to 
restrict to a hypersurface
in this cone as explained above.} It consists of two K\"ahler cones,
which are connected by a flop transition at $W=U$. Following 
\cite{Rahmfeld} we call the cones 
region III and region II, respectively. Inside region III the moduli
satisfy 
\begin{equation}
W >U>0 \mbox{   and   } T > W + \frac{1}{2} U
\label{region3}
\end{equation}
and the corresponding Calabi-Yau space is the above mentioned
elliptic fibration over ${\bf F}_1$. 
The prepotential is given by
\begin{equation}
{\cal V}_{III}(X)= \frac{5}{24} U^3 + \frac{1}{2} U T^2 
- \frac{1}{2} U W^2 + \frac{1}{2} U^2 W.
\end{equation}
This Calabi-Yau space is a $K_3$ fibration and therefore it
has a heterotic dual, which is obtained by compactifying the 
heterotic $E_8 \times E_8$ string on $K_3 \times S_1$ with instanton 
numbers $(13,11)$.
In addition to the flop boundary region III has two more boundaries:
at $W=T+\frac{1}{2}U$ the gauge symmetry is enhanced to $SU(2)$,
whereas one finds tensionless strings at $W=U$. 
Both boundaries are
boundaries of the extended cone, i.e. the moduli space ends at these
boundaries. 
\\
Passing through the flop transition at $U=W$ one enters region II, 
which is parametrized by
\begin{equation}
U>W>0 \mbox{   and   }T>\frac{3}{2} U.
\end{equation}
Here the prepotential takes the form \cite{Louis}
\begin{equation}
{\cal V}_{II}(X)= \frac{3}{8} U^3 + \frac{1}{2} U T^2 - \frac{1}{6} W^3
\end{equation}
The corresponding Calabi-Yau space is not a $K_3$ fibration and, therefore,
there is no dual (weakly coupled) heterotic description. 
Region II has two additional boundaries: 
The boundary $W=0$ corresponds to an elliptic
fibration over $P^2$. At the
other boundary $T=\frac{3}{2}U$ strings become tensionless. The 
effective action is regular everywhere in the extended K\"ahler cone
including those boundaries where no tensionless strings appear. 
\subsection{The Black Hole Solution in Region II \label{Sec2}}
In region II the dual (special) coordinates are given by 
\begin{eqnarray}
X_1={1\over 3} TU, \hspace{1cm}
X_2={1\over 6} T^2+{3\over8} U^2, \hspace{1cm}
X_3=- {1\over 6} W^2.
\end{eqnarray}
Let us denote the rescaled coordinates as, 
$Y^I = \cT, \cU, \cW$. The prepotential in terms of 
these quantities are given by,
\begin{equation}
\cV(Y) = Y_I Y^I = \frac{1}{6} C_{I J K} Y^I Y^J Y^K = \frac{3}{8}
{\cU}^3 + \frac{1}{2} \cU {\cT}^2 - \frac{1}{6} {\cW}^3
\end{equation}
The stabilization equations in terms of 
these rescaled coordinates are given by, 
\begin{eqnarray}
\cU \cT &=& H_{\cT} 
\nonumber\\
\frac{1}{2} (\cT)^2 + \frac{9}{8}(\cU)^2  
&=& H_{\cU}
\nonumber\\
- \frac{1}{2} (\cW)^2 &=& H_{\cW}
\end{eqnarray}
with solution
\begin{eqnarray}
\cT &=&  \sqrt{H_{\cU} (1 \pm \sqrt{1-\Delta})}
\nonumber\\
\cU &=& 
\frac{H_{\cT}}{\sqrt{H_{\cU} (1 \pm \sqrt{1-\Delta})}}
\nonumber\\
\cW &=& \sqrt{-2 H_{\cW}}
\label{SolReg2}
\end{eqnarray}
and $\Delta= \frac{9}{4} \frac{H_{\cT}^2}
{H_{\cU}^2}$.
Using the equivalent form
\begin{eqnarray}
\cU &=& \frac{2}{3} \sqrt{H_{\cU} (1 \mp \sqrt{1-\Delta})}.
\end{eqnarray}
one can check that the corresponding unrescaled scalar fields, 
when evaluated on the horizon, take the fixpoint values
found in \cite{Rahmfeld}.

Any additional negative sign appearing inside the roots of the
solutions yields an additional constraint in terms of 
charges/harmonic functions of the solution. 
In general the solution has to satisfy at least two bounds
\begin{eqnarray}
H_{\cW} \leq 0, \hspace{1cm}
H_{\cU}^2 \geq \frac{9}{4} H_{\cT}^2
\end{eqnarray}
Then it is manifest that $\cT, \cU, \cW > 0$. The condition 
for the solution to be inside the K\"ahler cone is given by, 
$\frac{2}{3} \cT > \cU > \cW > 0$. For the condition $\cU > \cW$, 
we get
\begin{equation}
H_{\cW}^2 + \frac{4}{9} H_{\cU} H_{\cW} + \frac{1}{9} H_{\cT}^2 > 0
\end{equation}
At the 
boundary, $\cU = \cW$, so the condition for reaching the boundary is,
\begin{equation}
H_{\cW} = - \frac{2}{9} H_{\cU} \pm \frac{1}{3} \sqrt{\frac{4}{9}
H_{\cU}^2 - H_{\cT}^2}
\end{equation}
If $1 - \Delta \geq 0$, then we can have at least one solution 
with $H_{\cW} < 0$. This shows that the solution can cross the K\"ahler 
cone while still respecting the other constraints. 

Next consider the condition $\cT > \frac{3}{2} \cU$. In terms of 
harmonic functions, we get
\begin{equation}
H_{\cU} + \sqrt{H_{\cU}^2 - \frac{9}{4} H_{\cT}^2} > \frac{3}{2} 
H_{\cT}
\end{equation}
This is obviously satisfied if the term under the square root 
is positive. Note that this time at the boundary, the moduli of
$\cT, \cU$ would become imaginary when further continuing the 
solution. The behaviour at this boundary is different from that 
at the flop boundary and one does not expect to be able to extend 
the solution. Analogous remarks apply to the boundary $\cW = 0$. 

The corresponding ADM mass of the black hole solution 
can be determined by the central charge evaluated at spatial infinity,
i.e. using 
$Z_\infty = q_I Y_\infty^I = q_I X_\infty^I$ one obtains
for our particular black hole solution
\begin{equation}
Z_\infty =     
  q_{\cT} \sqrt{h_{\cU}} \ 
 \left [
   \sqrt{(1 \pm \sqrt{1 - \delta (h)})} 
   + \frac{1}{\sqrt{(1 \pm \sqrt{1 - \delta (h)})} }
 \right ] \ + \ q_{\cW} \sqrt{-2 h_{\cW}}
\end{equation}
with
\begin{equation}
 \Delta (r \rightarrow \infty) \equiv \delta (h) = \frac{9 h_{\cT}^2}
{4 h_{\cU}^2}.
\end{equation}
The entropy of the black hole solution is given by the
central charge of the $N=2$ supersymmetry algebra evaluated at the
horizon $r=0$:
\begin{equation}
Z_h = \lim_{r \rightarrow 0} 3^{1/3} \
\left (   
  r q_I Y^I 
\right )^{2/3}
\end{equation}
Our solutions at the horizon are, 
\begin{eqnarray}
\lim_{r \rightarrow 0}{r \cT} &=& \sqrt{q_{\cU} \pm 
\sqrt{q_{\cU}^2 - \frac{9}{4} q_{\cT}^2}} 
\nonumber \\
\lim_{r \rightarrow 0}{r \cU} &=& \frac{q_{\cT}}{\sqrt{q_{\cU} \pm \sqrt
{q_{\cU}^2 - \frac{9}{4}q_{\cT}^2}}} 
\nonumber \\
\lim_{r \rightarrow 0}{r \cW} &=& \sqrt{- 2 q_{\cW}}
\end{eqnarray}
Thus, in region II one obtains 
\begin{equation}
Z_h =  3^{1/3} \
\left (   
  q_{\cT} \sqrt{q_{\cU}} \ 
 \left [
   \sqrt{(1 \pm \sqrt{1-\delta (q)})} 
   + \frac{1}{\sqrt{(1 \pm \sqrt{1-\delta (q)})} }
 \right ] \ - \frac{1}{2} (-2 q_{\cW})^{3/2}
\right )^{2/3}
\end{equation}
with
\begin{equation}
 \Delta_h \equiv \delta (q) = \frac{9 q_{\cT}^2}{4 q_{\cU}^2},
\end{equation}
which can be checked to coincide with the result of
\cite{Rahmfeld} by extremization of the central charge.
\subsection{The Black Hole Solution in Region III \label{Sec3}}
In region III a field identification of the form
\begin{eqnarray}
W = S^\prime - \frac{1}{2} (T^\prime - U^\prime), 
\hspace{1cm}
T = S^\prime + \frac{1}{2} T^\prime,
\hspace{1cm}
U = U^\prime
\end{eqnarray}
yields the prepotential 
\begin{equation}
{\cal V}_{III}(X)= S^\prime T^\prime U^\prime + \frac{1}{3} U^{\prime 3}. 
\end{equation}
The corresponding black hole solution of this prepotential within a
different model\footnote{The fact that the prepotentials can be
brought to the same form does not apply that the K\"ahler 
cones are the same. See \cite{Louis} for a 
detailed discussion.}
has been already studied in \cite{Gaida}.
Since we can express all the fields in both chambers
using special coordinates ($T,U,W$), 
we keep this parametrization of the prepotential which is given 
by,
\begin{equation}
\cV = \frac{5}{24} U^3 + \frac{1}{2} U T^2 - \frac{1}{2} U W^2 
+ \frac{1}{2} U^2 W
\end{equation}

It follows that 
the dual coordinates in region III are given by 
\begin{eqnarray}
X_1&=&{1\over 3} TU
\nonumber\\
X_2&=&{1\over 6} T^2-{1 \over 6} W^2 + {1\over 3} UW
      + {5 \over 24} U^2
\nonumber\\
X_3&=&- {1\over 3} UW  + {1 \over 6} U^2 
\end{eqnarray}
Again in terms of rescaled coordinates, the stabilization equations read
\begin{eqnarray}
\cT \cU &=& H_{\cT}
\nonumber\\
(\cT)^2 - (\cW)^2 + 2 \cU \cW + \frac{5}{4}(\cU)^2  
&=& 2 \ H_{\cU}
\nonumber\\
(\cU)^2 - 2 \cU \cW &=& 2 \ H_{\cW}
\end{eqnarray}
One obtains the following set of solutions
\begin{eqnarray}
\cT &=& 
 \frac{H_{\cT}}{\sqrt{ a \pm \sqrt{a^2 + b}}}
\nonumber\\
\cU &=& \sqrt{ a \pm \sqrt{a^2 + b}}
\nonumber\\
\cW &=&  - \frac{H_{\cW}}{\sqrt{ a \pm \sqrt{a^2 + b}}}
+ \frac{1}{2} \sqrt{ a \pm \sqrt{a^2 + b}}
\label{SolReg3}
\end{eqnarray}
with
\begin{eqnarray}
a = \frac{1}{2} (H_{\cU} + \frac{1}{2} H_{\cW}),
\hspace{2cm}
b = \frac{1}{2} (H_{\cW}^2 -  H_{\cT}^2)
\end{eqnarray}
Again additional negative signs appearing in the roots of the
solutions yield additional constraints. The reality of the moduli 
gives the constraint,
\begin{eqnarray}
a^2 + b \geq 0.
\end{eqnarray}
In terms of harmonic functions, this condition reads as,
\begin{equation}
9 H_{\cW}^2 + 4 H_{\cU}^2 - 8 H_{\cT}^2 + 4 H_{\cW} H_{\cU} 
\geq 0
\end{equation}
{From} the solutions, $\cU \geq 0$ is automatically satisfied. 
Analysis of $\cW \geq 
\cU$ implies that $H_{\cW} < 0$. Assuming $H_{\cU} > 0$ and 
$- 9 H_{\cW} - 2 H_{\cU} > 0$, the constraint is obtained as
\begin{equation}
9 H_{\cW}^2 + 4 H_{\cW} H_{\cU} + H_{\cT}^2 \geq 0
\end{equation}
Finally let us consider $\cT > \frac{\cU}{2} + \cW$.  
It is useful to analyze the simpler and stronger constraint 
$\cT \geq \frac{3}{2}\cU$, which will be relevant while constructing
solutions interpolating between chambers II and III. Assuming
$8 H_{\cT} - 3 H_{\cW} - 6 H_{\cU} > 0$, this implies,
\begin{equation}
17 H_{\cT}^2 - 9 H_{\cW}^2 - 6 H_{\cT}H_{\cW} - 12 H_{\cT}H_{\cU}
\geq 0
\end{equation} 
The ADM mass of the solution is determined by 
\begin{eqnarray}
Z_\infty &=& \frac{1}{\cU_\infty} (q_{\cT} h_{\cT} - q_{\cW} h_{\cW})
  + \cU_\infty (q_{\cU} + \frac{1}{2} q_{\cW}).
\end{eqnarray}
Again one can compute the black hole entropy from the central charge
evaluated at the horizon at $r=0$. The result is 
\begin{equation}
Z_h =  3^{1/3} 
\left (   
 \frac{8 q_{\cT}^2 + 4 q_{\cU} q_{\cW} + 4 q_{\cU}^2 - 
7 q_{\cW}^2 \pm (q_{\cW} + 2 q_{\cU})
\alpha (q)}
{4 \sqrt{q_{\cW} + 2 q_{\cU}  \pm \alpha (q)} }
\right )^{2/3}
\end{equation}
with
\begin{eqnarray}
\alpha^2 (q) &=& 9 q_{\cW}^2 + 4 q_{\cW} q_{\cU} + 
4 q_{\cU}^2 - 8 q_{\cT}^2
\end{eqnarray}

\subsection{Interpolating solutions \label{SecInter}}
In this section we present an explicit example of
a massive and regular supersymmetric black hole solution,
where the scalar fields take values in region II at
infinity but values in region III at the horizon.
Thus the solution interpolates between the two $N=2$ vacua,
namely, the supersymmetric 
'flat space $\times$ $ CY_{II}$' at $r=\infty$ and 
'horizon $\times $ $CY_{III}$' at $r=0$. The geometry
of the horizon and the restoration of full $N=2$ supersymmetry
on it was found in \cite{CFGK}. Our point here is that one can
take  $CY_{II}$ and $CY_{III}$ to be topologically
distinct Calabi-Yau spaces associated with the regions
II and III of the extended K\"ahler cone. Consequently
the solution crosses the flop line $U=W$ in moduli space
for some specific value $r=r^{\star}$ of the radius. 
We will show explicitly that one can glue appropriate solutions
of region II and region III such that
the scalar fields are continuous but not smooth (i.e. not $C^{\infty}$)
at $r=r^{\star}$.

The equations of motions of scalars coupled to gravity
can be cast in the form of a generalized geodesic equation
mapping space-time to moduli space. In the case of static
supersymmetric four- and fivedimensional black holes
this equation shows a fix point behaviour \cite{FerKal1,CFGK}.
This means that  the values of the scalar fields at the horizon are
uniquely fixed by the charges. In contradistinction, 
the values of the scalars at infinity can be changed 
continuously. The equations of motion determine the
flow from the values specified at infinity to their fixpoint values
at the horizon.

We will now make a specific choice for the parameters 
$h_{i}$ and $q_{i}$, which control the asymptotic behaviour at 
$r=\infty$ and $r=0$, respectively. Our strategy will be
to impose that the solution is in the interior of 
region II at infinity, but in the interior of region III
at the horizon. Beside this we will make 
the explicit expressions we get at $r=r^{\star}$ as
simple as possible, so that they allow for an exact analytical 
treatment. Note that with a generic choice of parameters one
is likely to get higher order algebraic equations as 
matching conditions at $r=r^{\star}$, which need not have
a general solution at all. We would, however, like to have an
explicit example where we can compute analytically the 
behaviour of
the scalar fields and of the space-time metric at the 
transition point $r=r^{\star}$. Our example is
otherwise generic, because the scalar fields take 
values on a special line in moduli space only at a
single radius $r=r^{\star}$. Therefore we can expect that the
qualitative behaviour exhibited in the example is generic.

In order to be in region II at infinity we need to impose 
$\cU_{\infty} > \cW_{\infty} > 0$ and
$\cT_{\infty} > \frac{3}{2} U_{\infty}$.
In addition we have
to impose $\cV(Y)_{\infty}=1$ in order to have an
asymptotically  flat configuration. Using the solution
(\ref{SolReg2})
in region II, 
we convert these conditions into restrictions on the asymptotic
values $\hW,\hU,\hT$ of the harmonic functions.\footnote{We
take the '$+$' branch in (\ref{SolReg2}).} 
We will assume that $\hW<0$, $\hU,\hT>0$. 
Then being in region II implies 
\be
\hU > \frac{3}{2} \hT,\;\;\;
\hT^{2}> - 4 \hW \hU.
\label{AsymIn2}
\ee

Next we have to impose that the scalars at $r=0$
satisfy (\ref{region3}), i.e. that they are in region III
at the horizon. In terms
of our rescaled variables this becomes
\be
(r\cW)_{r=0} > (r \cU)_{r=0} > 0 
\mbox{   and   }
(r\cT)_{r=0} > (r\cW)_{r=0} + \frac{1}{2} (r \cU)_{r=0}.
\label{ChargeIn3}
\ee
The constraints resulting from
(\ref{region3}) have already been
discussed in section \ref{Sec3}. Here we will simply
specify a set of charges which is easily seen to satisfy
(\ref{ChargeIn3}).

It turns out that the following choice of charges
and asymptotic parameters is convenient:
\be
\begin{array}{lll}
\hW = - \frac{1}{18} h, & \hU = \frac{5}{2} h, & \hT = h,\\
\qW = - \frac{2}{9} q, & \qU = \frac{5}{2} q, & \qT = q, \\
\end{array}
\label{parameters}
\ee
with $h>0$ and $q>0$. This satisfies both 
(\ref{AsymIn2}) and (\ref{ChargeIn3}). Now $h$ is not a free
parameter but fixed by demanding asymptotic flatness
${\cal V}(Y)_{\infty}=1$. We will not
need the explicit value of $h$, but evaluation of $\cV(Y)_{\infty}=1$
shows that it is positive, and therefore the set of parameters
(\ref{parameters}) is consistent. The advantage of this 
choice is that it drastically simplifies the solution
in region II 
because $H_{\cU} = \frac{5}{2} H_{\cT} = \frac{5}{2} H$
and therefore 
\be
\cW = \sqrt{-2 H_{\cW}}, \;\;\;
\cT = \sqrt{ \frac{9}{2} H}, \;\;\;
\cU = \sqrt{ \frac{2}{9} H}. 
\ee
for $r>r^{\star}$.

Since the charges are chosen such that the solution
is in region III at the horizon we expect to find
$\cW(r^{\star})=\cU(r^{\star})$ for some value $r=r^{\star}$
of the radius. 
For the above choice of data one indeed finds
$r^{\star} = \sqrt{2 \frac{q}{h}}$ with
\be
\cW(r^{\star}) = \frac{\sqrt{3}}{3} \sqrt{h} =
\cU(r^{\star}), \;\;\;
\cT(r^{\star}) = \frac{3}{2} \sqrt{3} \sqrt{h}.
\label{boundary}
\ee
For $r<r^{\star}$ the scalar fields further evolve
according to the region III solution to the fix point at $r=0$.
Thus we need to specify a solution (\ref{SolReg3})
in region III,
such that it satisfies the boundary conditions
(\ref{boundary}) and reaches the fixpoint.
Such a solution depends on a second set of harmonic
functions, $H_{i}'=h_{i}' + \frac{q_{i}}{r^{2}}$.
Note that the charges appearing in the 
functions $H_{i}'$ have to be the same as in the $H_{i}$, 
since we have already chosen our fixpoint.
The asymptotic constants $h_{i}'$ 
have to be determined from (\ref{boundary}).
One finds that no matching is possible if one
takes the '$+$' sign in (\ref{SolReg3}) whereas with
the choice of the '$-$' in (\ref{SolReg3}) the matching
condition (\ref{boundary})
is satisfied if and only if $h_{i}' = h_{i}$.
Thus we conclude that the harmonic functions
$H_{i}'$ and $H_{i}$ have to be 
the same in both regions.  
This yields a unique continuation of the
region II solution through the flop transition into
region III, which reaches the fixpoint specified by the
charges.

We can now analyze how the scalar fields, 
the prepotential $\cV(Y)$ and the space-time
metric behave at the flop transtion.\footnote{The results
reported in the following have been obtained using Maple.}
The result is that
scalar fields are $C^{1}$-functions with respect to the radius $r$, 
i.e. they are continuously differentiable. The second derivatives of the
scalar fields are not continuous at $r=r^{\star}$ but
jump by a finite amount. In the prepotential 
$\cV(Y)$, which is a cubic polynomial in the scalars and which 
determines the space-time metric the discontinuities in the second
derivatives precisely cancel. However the third derivative 
is discontinuous at the flop, i.e. $\cV(Y)$  
is a $C^{2}$-function. 
As a consequence the space-time Riemann tensor is continuous 
at the flop transition, because it depends on no higher
derivative of the metric then the second. One also expects
that the Riemann tensor is not continuously differentiable
at the flop, because of the discontinuity of the third derivative
of $\cV(Y)$. But since the Riemann tensor depends on $\cV(Y)$ and
its derivatives in a complicated way, one could imagine a cancelation
of the discontinuities, as we have seen happening for $\cV(Y)$ itself.
However one can check by explicit computation 
that the Riemann tensor is continuous, but not differentiable in $r$
at the transition point. Namely, the $r$-derivatives of
those components of the Riemann tensor containing the
index $r$ jump by a finite amount at $r=r^{\star}$.

This shows that the space-time geometry is regular, but
not smooth at the transition. The phase transition
in the internal Calabi-Yau space manifests itself in 
a 'roughening' of the space-time geometry.


\subsection{Massless black holes at the flop transition \label{SecMassless}}
In this subsection we analyse the conditions to obtain
massless black holes at the flop transition.
\\
The relevant quantity to consider massless black holes
is of course the ADM mass or, equivalently, the central charge
evaluated at spatial infinity. 
For our particular model the latter reads in general
\begin{equation}
Z_{\infty} = q_{\cT} {\cT}_{\infty} + q_{\cU} {\cU}_{\infty} 
+ q_{\cW} {\cW}_{\infty}
\end{equation}
At the flop transition the solution has to satisfy
${\cU}_{\infty} = {\cW}_{\infty} \geq 0$. Thus,
at the flop transition the first constraint on the moduli at spatial
infinity is given by
\begin{equation}
\frac{2}{3} {\cT}_{\infty} > {\cU}_{\infty} = {\cW}_{\infty} \geq 0
\end{equation}
and restricts the parameters $h_I$, so that the moduli lie in
the physical moduli space.
It follows that one of the three parameters $h_I$ can be eliminated
at the flop transition.
Moreover one obtains from the stabilisation equations
\begin{equation}
{\cW}_{\infty} = {\cU}_{\infty} = \sqrt{- 2 h_{\cW}}, \hspace{1cm}
{\cT}_{\infty} = \frac{h_{\cT}}{\sqrt{- 2 h_{\cW}}}.
\end{equation}
{From} the first constraint it follows that
\begin{equation}
h_{\cT} > -3 h_{\cW} > 0
\end{equation}
must hold. On the other hand we have an additional constraint, which
must hold at spatial infinity, namely
\begin{equation}
{\cV}_{\infty} \ = \ 1 \ = \ \frac{5}{24} {\cW}^3_{\infty}
              + \frac{1}{2} {\cW}_{\infty} {\cT}^2_{\infty}. 
\end{equation}
This second constraint yields
\begin{equation}
h_{\cT}^2 =  2 (-2 h_{\cW})^{1/2} - \frac{5}{12} (- 2 h_{\cW})^2.
\end{equation}
Thus, the solution at the flop transition depends in general
on three charges and one parameter, say $h_{\cW}$, only.
{From} the first constraint follows now, that the parameter
$h_{\cW}$ is constrained, i.e.
\begin{equation}
4 (-2 h_{\cW})^{3/2} \leq 3.
\end{equation}
This means that $|h_{\cW}|$ has to be sufficiently small, such that
the set of parameters denotes a valid black hole solution. If the
solution does not satisfy this constraint, 
then it is either not in the physical moduli space or 
can not be asymptotically flat. 

In order to obtain massless black holes we have to analyze the
condition $Z_{\infty}=0$. Writing this out in terms of independent
parameters yields
\be
\qT \sqrt{ 2 \sqrt{-2 \hW} - \frac{5}{12} (-2 \hW)^2 }
+ (\qU + \qW) (-2 \hW) = 0
\label{vanishingZ}
\ee
As explained in \cite{Witten3} consistency of the model requires that
there is an extra non-perturbative state present at the line
$U=W$. In order to identify this state with a massless black hole we 
need to have $Z_{\infty}=0$ for the whole line $U=W$, that is for
all values of $\hW$. This implies $\qT=0$ and $\qU + \qW=0$ and fits
with the geometric picture of a M-theory two-brane wrapped on a vanishing
two-cycle. As explained in \cite{Louis} $U-W$ is the modulus
associated with this vanishing cycle. Wrapped two-branes with $\qT=0$ and
$\qU + \qW=0$ wind  around the cycle whose size is measured by 
$U-W$, but do not wind around any other two-cycle.
Therefore precisely these winding states
become massless at the flop transition.\footnote{In order
to have a single massless multiplet one has to assume that
multiple winding states are multiple particle states, rather then 
multiply-charged single particle states. This is hard
to verify directly, because the multiple winding states are bound
states at threshold \cite{Rahmfeld}.}
We conclude that the spectrum of massless black holes 
agrees with the spectrum of massless non-perturbative states predicted
by M-theory. Note however that the charge configuration one has
to take is such that the solution at the horizon is not inside the
extended K\"ahler cone. This means that the solution meets the
boundary at finite $r>0$. Thus, the term massless black hole might
be a misnomer. Clearly the behaviour of such solutions deserves 
further study.

One might also wonder whether there are additional massless black holes
corresponding to zeros of $Z_{\infty}$ at special points within the
line $U=W$. Such multicritical points would correspond to more
complicated singularities of the Calabi-Yau space, occuring in
higher codimension in moduli space. Assuming $\qT \not= 0$
one can solve for $\hW$ with the result:
\begin{equation}
h_{\cW}^3 = - \frac{1}{2} 
\left (
 \left (
  \frac{q_{\cW} + q_{\cU}}{q_{\cT}}
 \right )^2
 + \frac{5}{12}
\right )^{-2}  
\end{equation}
In addition these charges must
satisfy all the constraints given above. 
It would be very interesting to find the explicit quantum numbers for
an additional massless black hole at the flop transition. However,
because of the number of constraints and the 
complicated form of the black hole
solution, it seems that there is no straightforward analytical
way to find such a solution. In addition a possible relation to 
higher degenerations of the Calabi-Yau space is not clear to us.
These questions are beyond the scope of this article and we leave them
for future investigations.   

\section{Summary \label{SecCon}}
 
In this paper we have presented generic extreme black hole solutions
of M-theory compactified on a specific family of Calabi-Yau threefolds.
In particular their ADM masses and BH entropies were computed. Our 
results contain those obtained from studying double extreme solutions
through extremization of the central charge in \cite{Rahmfeld} as 
a subset and they further confirm the attractor picture \cite{FerKal1}
of BPS black holes. The picture that we get 
of phase transitions and
massless black holes nicely fits with the one found in four dimensions
\cite{BehLueSab}. 

In particular we showed by an explicit example that
black hole solutions can interpolate between topologically
distinct (though birationally equivalent) internal Calabi-Yau spaces.
Despite the fact that the geometry of the internal space is truly singular
at a flop transition, this is reflected in the space-time geometry
only by a ``roughening'', namely a discontinuity in the radial derivative
of certain components of the Riemann tensor. We also showed that
one can obtain massless solutions at the flop transition line, 
which can be identified with the extra massless states
predicted by M-theory \cite{Witten3}. However these solutions
have charge configurations which force the solution to run into the
boundary of the extended K\"ahler cone before a horizon is reached.
Clearly these massless solutions, as well as those related
to the other boundaries of the K\"ahler cone, need further study.


{\bf Acknowledgments}

I.G. would like to thank the Department of Physics at 
Martin-Luther-University for its warm hospitality.
S.M. gratefully acknowledges the Abdus Salam International Centre
for Theoretical Physics for an Associateship, under which a part  
of this work has been done.

\medskip



\end{document}